\documentclass{elsart}
\usepackage{amsmath}
\usepackage{amssymb}
\usepackage{graphics}
\begin{document}

\begin{frontmatter}
\title{Low-Energy D-Wave Effects in Neutral Pion Photoproduction}

\author{C. Fern\'andez-Ram\'{\i}rez},
\ead{cefera@mit.edu}
\author{A.M. Bernstein},
\author{T.W. Donnelly}

\address{Center for Theoretical Physics,
Laboratory for Nuclear Science and Department of Physics, Massachusetts Institute of Technology,
77 Massachusetts Ave., Cambridge, Massachusetts 02139, USA}

\begin{abstract}
The contribution of D waves to physical observables for neutral
pion photoproduction in the near threshold region is studied. 
Heavy Baryon Chiral Perturbation Theory to one loop, and up to $\mathcal{ O}(q^4)$,
is used
to account for the S and P waves, while D waves are added 
in an almost model-independent way using standard Born
terms and vector mesons.
It is found that the inclusion of D waves is necessary
to extract the $E_{0+}$ multipole reliably from present and forthcoming data
and to assess the low-energy constants of Chiral Perturbation  Theory.
Arguments are presented demonstrating that F-wave contributions are negligible
in the near-threshold region.
\end{abstract}

\end{frontmatter}

Due to the spontaneous breaking of chiral symmetry in Quantum Chromo Dynamics (QCD) the $\pi$ meson appears as a pseudoscalar Nambu-Goldstone Boson \cite{book}. The dynamic consequences are that
the S-wave amplitude for the $\gamma N \rightarrow \pi^{0} N$ reaction is small in the 
threshold region, since it vanishes in the chiral limit, \textit{i.e.} when the light quark masses 
are set equal to zero 
\cite{CHPT91,CHPT96,CHPT01}. An additional  consequence is that the 
P-wave amplitude is large and leads to the $\Delta$ resonance at intermediate
energies \cite{AB-Delta}.
Accordingly  the photoproduction of neutral pions differs from the general pattern 
for hadronic reactions where
the S wave dominates close to threshold and then, as the energy increases, the higher angular momentum states (P, D, \ldots)  start to become important. By contrast, for the 
$\gamma N \rightarrow \pi^{0} N$ reaction the S- and P-wave contributions
are comparable  even very close to threshold \cite{AB-fits}.

The purpose of this Letter is to show for the first time that the D waves also play a significant role in the near-threshold region (up to a photon energy $\sim$165 MeV) and to this end several observables have been extended 
beyond the traditional S+P wave limit
to include the D-wave contributions. 
Fits have been performed for the latest and most accurate $\gamma p \rightarrow \pi^{0} p$
data in the near-threshold region \cite{Schmidt},
in contrast to all previous analyses for which the data have been described using only S and P waves 
\cite{CHPT91,CHPT96,CHPT01,Schmidt,AB-fits}. It  will be shown that this affects the value of the extracted S-wave amplitude ($E_{0+}$), particularly its energy dependence. In contrast the P-wave multipoles are not affected by the inclusion of  the D waves. 

Specifically, we explore the impact of D-wave contributions in the near-threshold region
using as starting point Heavy Baryon Chiral Perturbation Theory (HBCHPT) \cite{CHPT96,CHPT01}.
We compute the S and P waves employing HBCHPT to one loop, and up to $\mathcal{ O}(q^4)$, 
and add the 
D waves using standard Born terms (equivalent to the Born contribution to HBCHPT)
and vector mesons \cite{FMU06}. The Low Energy Constants (LECs) of 
HBCHPT are assessed through fits to the latest experimental data from Mainz \cite{Schmidt}
and the multipoles extracted.

Let us begin by summarizing the basic structure of the observables and multipoles.
It is always possible to expand the differential cross section in terms of Legendre polynomials
\begin{equation}
\sigma_T (\theta)= \frac{q_\pi}{q_\gamma} \left[
T_0 + T_1 \mathcal{P}_1\left( \theta \right)  
+ T_2 \mathcal{P}_2\left( \theta \right) + \dots \right] \: , \label{eq:sigma}
\end{equation}
where $q_\pi$ and $q_\gamma$ are the pion and photon momenta in the center-of-mass respectively
and the $T$s depend on the photon energy.
Following the developments in \cite{DonnellyRaskin}
we have worked out the 
full multipolar expansion of the $T$s up to D waves.
This expansion and a full analysis of near-threshold neutral pion photoproduction
will be detailed in future work \cite{FBD09}. 
If we take into account only S and P waves, then only $T_0$, $T_1$, and $T_2$ coefficients contribute. 
If we add D waves,
two more quantities appear, $T_3$ and $T_4$.
So the first place to look for D waves consists in checking to see if there is room for the appearance
of these new terms. The currently available experimental data \cite{Schmidt}
can be described quite well using only $T_0$, $T_1$, and $T_2$, 
and no $T_3$ or $T_4$ contribution appears to be required at present.
Hence, any contribution of D waves to $\sigma_T$ should appear as a modification of
$T_0$, $T_1$ or $T_2$. On the other hand, $T_0$ and $T_2$ are dominated by diagonal terms
involving the multipoles, namely 
$|M_{1+}|^2$, $|M_{1-}|^2$, and $|E_{0+}|^2$, and thus any interference with D waves would be negligible compared to the leading terms.
On the other hand, $T_1$ is entirely due to multipole interferences:
$T_1$ up to D waves can be written:
\begin{equation}
T_1= 2 \: \text{Re} \left[ P_1^* E_{0+} \right] + \delta T_1 \: , \label{eq:t1}
\end{equation}
where $P_1\equiv 3E_{1+}+M_{1+}-M_{1-}$, 
and $\delta T_1$ stands for the D-wave/P-wave interference contribution
\begin{equation}
\begin{split}
\delta T_1 &= 2 \: \text{Re} \Big{[} \: \frac{27}{5}M^*_{1+}M_{2+} 
+\left( M^*_{1+} - M^*_{1-} \right) E_{2-} \\
&+E^*_{1+} \left( \frac{72}{5}E_{2+}-\frac{3}{5}E_{2-} +\frac{9}{5}M_{2+}
-\frac{9}{5}M_{2-} \right) \\
&+ \left( \frac{3}{5}M^*_{1+}+3M^*_{1-} \right) M_{2-}   \Big{]} \: .
\end{split}
\end{equation}
The P waves interfere with the D waves, enhancing their influence in the observable and
possibly
compromising any multipolar extraction from data.
Note that there is no interference contribution to $\delta T_1$ between $E_{0+}$ and the higher
partial waves.

To account for the S and P waves, 
the best available theoretical framework to study pion photoproduction in the near threshold region
is HBCHPT. In \cite{CHPT96,CHPT01} the S and P multipoles to one loop
and up to $\mathcal{ O}(q^4)$ are provided and we take these as our starting point. Six LECs appear,
and five have been fitted to the data \cite{Schmidt}:
$a_1$ and $a_2$ associated with the $E_{0+}$ counter-term, $b_p$ associated with the
$P_3 \equiv 2M_{1+}+M_{1-}$ multipole, together with
 $\xi_1$ and $\xi_2$ associated with $P_1$ and 
$P_2 \equiv 3E_{1+}-M_{1+}+M_{1-}$, respectively.
The $c_4$ LEC associated with $P_1$, $P_2$, and $P_3$
has been taken from \cite{MeissnerNPA00}
where it was determined from pion-nucleon scattering inside
the Mandelstam triangle.
The corresponding details and fixed parameters, such as the pion-nucleon coupling 
$g_{\pi N}$,  can be found in \cite{CHPT96,CHPT01}.
We obtain the D waves 
from the customary Born terms and vector meson exchange ($\omega$ and $\rho$) 
\cite{FMU06}.
For the $\omega$ and $\rho$ parameters we have used
those given by the dispersion analysis of the form factors in \cite{MMD}.
The vector-meson correction is very small and the inclusion of D waves in this way is almost 
equivalent to the zeroth order in HBCHPT.

\begin{figure}
\rotatebox{-90}{\scalebox{0.33}[0.33]{\includegraphics{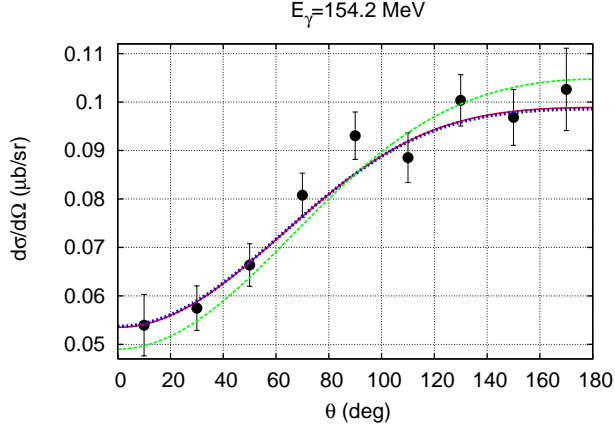}}}
\caption{(Color online) Differential cross section at $E_\gamma=154.2$ MeV. 
The experimental points are from \cite{Schmidt}. The errors are statistical and do not
include the 5\% systematic error.
We display results using the following models,
SP (dotted), SPD (solid),
and SP with D waves added without refitting the data (dashed).
The SP and SPD curves mostly overlap.} \label{fig:dsigma}
\end{figure}

Within this framework we have performed two fits
to the latest experimental data from Mainz \cite{Schmidt} 
(171 differential cross sections and 7 photon asymmetries, spanning the energy range from threshold up to 166 MeV), one including only S and P waves (SP model) that was already done in \cite{CHPT01}, and another where we have added the D waves (SPD model).
Both fits have the same number of parameters, which are the named LECs in HBCHPT.
We obtain $\chi^2/dof=1.23$ for the SP model and $\chi^2/dof=1.25$ for the SPD model.
The 70\% (90\%) confidence level is set to $\chi^2/dof=1.27 (1.29)$ and $\chi^2/dof=1.28 (1.30)$, respectively.
The P waves and their related LECs 
prove to be very stable against the inclusion of D waves,
and indeed they are almost the same for both the SP and SPD fits and those given in \cite{CHPT01},
with differences lying typically below the 1\% level.
Of special interest is $P_1$ which has $\xi_1$ as its sole free LEC and is the one that could compromise the $E_{0+}$ extraction.
Both the SP and SPD models provide equally good fits to data from \cite{Schmidt}. 
The P-wave parameters $b_p$, $\xi_1$, and $\xi_2$
happen to be larger than those expected from
resonance saturation. For an extensive analysis of this issue we refer the reader to \cite{CHPT01}.
The parameters of our SP fit differ from the ones in \cite{CHPT01} and we obtain a better $\chi^2/dof$. This is due to our better optimization procedure. We use a genetic algorithm combined with a standard
gradient based routine, which provides us with a very powerful optimization tool. The details on the algorithm and its advantages can be found in \cite{PRC08}.
In Fig. \ref{fig:dsigma} we provide an
example of how well the fits compare with experimental data from \cite{Schmidt}. 
In addition we also provide the effects
of D waves when added to the SP model without refitting.

\begin{figure}
\rotatebox{-90}{\scalebox{0.33}[0.33]{\includegraphics{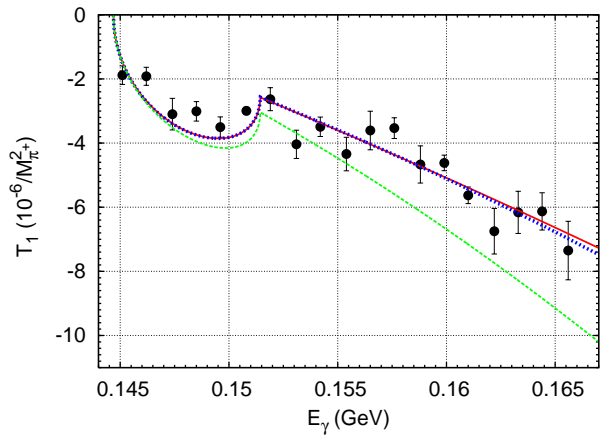}}}
\caption{(Color online) $T_1$ extracted  
from data \cite{Schmidt} compared with results using the following models, 
SP (dotted), SPD (solid),
and SP with D waves added without refitting the data (dashed). When theory is fitted to data including the D waves, the $E_{0+}$ multipole is modified to compensate for the D-wave contribution.
The SP and SPD curves mostly overlap.} \label{fig:T1}
\end{figure}

To obtain the \textit{experimental} $T_1$  displayed in Fig. \ref{fig:T1} we have fitted each energy data set from \cite{Schmidt} to Eq. (\ref{eq:sigma}) using the least-squares method and assuming that only $T_0$, $T_1$, and $T_2$ contribute. In this way, we can extract $T_0$, $T_1$, and $T_2$ in an almost model-independent way (the only assumption being that higher-order $T$s are negligible, as is the case
given the current measurements). Together with the \textit{data}, in Fig. \ref{fig:T1} we
plot $T_1$ for the SP and SPD fits. They both fit the same data from \cite{Schmidt}
and so they provide a very similar $T_1$, although the multipolar contributions differ. As an exercise we also plot the SP model adding the D waves
to explore how much they contribute to the observable. The modification of $T_1$ is meaningful, and hence the SPD model becomes necessary. As a consequence of the stability of the P waves, the $E_{0+}$ has to be modified to compensate the
D waves which pull $T_1$ to more negative values.
The  curves show the influence of the unitary cusp in $\text{Re} E_{0+}$, and are consistent with the data. To determine the magnitude of the cusp more accurately 
will require experiments with polarized targets \cite{AB-review}. 

\begin{figure}
\rotatebox{-90}{\scalebox{0.33}[0.33]{\includegraphics{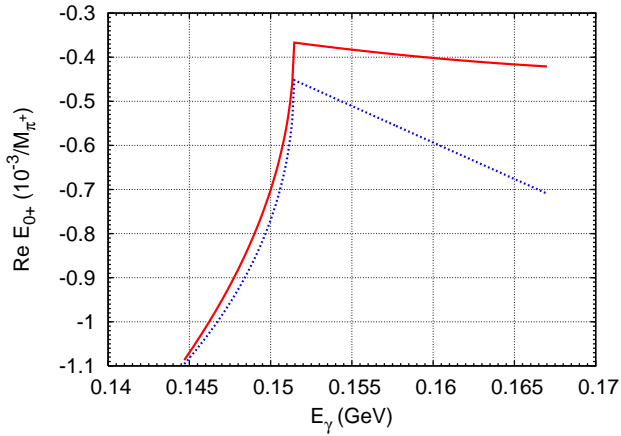}}}
\caption{(Color online) Extracted $E_{0+}$ multipole. The solid line corresponds to the SPD model and the dotted line to the SP model.} \label{fig:e0+}
\end{figure}

When we analyze the fits, we find that P-wave multipoles are very stable against the inclusion of the D waves. However, $E_{0+}$ is significantly changed through a modification of the $a_1$ and $a_2$ parameters, \textit{i.e.} via their impact on the counter-term (see below).
Hence, the absence of D waves in the analysis compromises the extraction of the $E_{0+}$ multipole
and its associated LECs.
In Fig. \ref{fig:e0+} we compare the two extractions of the $E_{0+}$ multipole. We can see that the
slope and the unitary cusp are dramatically changed by the inclusion of the D waves,
pointing out that D-wave contribution is meaningful even at the $\pi^+$ threshold.

Regarding which D-wave interferences play the major role in the modification of $T_1$, 
we have found that
these are the interferences firstly of $M_{2+}$ and secondly of $M_{2-}$ with the dominant
multipole $M_{1+}$.
At this point one might wonder if F waves could also make an important contribution to $T_1$
in this energy region through their interference with P waves.
The answer is no because, due to symmetry considerations, there is no interference of F and P waves
in $T_1$. Indeed, F waves only interfere with D and G waves, guaranteeing a negligible contribution
to $T_1$ in the near-threshold region.
The argument is as follows: first, for $T_1$ only partial waves with opposite parity can interfere
(see Eq. (\ref{eq:t1}) ); second, these partial waves can only differ by one unit of angular momentum,
unlike $T_i$ with $i>1$. Thus, one obtains the structure
$T_1= S \times P + P \times D + D \times F + F \times G + G \times H + \dots$

If we compare the SP and SPD fits for $E_{0+}$ we find that only two parameters can be modified,
and those are $a_1$ and $a_2$, the LECs that appear in the $E_{0+}$ counter-term:
\begin{equation}
E_{0+}^{ct}=ea_1 \omega M_\pi^2+e a_2 \omega^3 \: , \label{eq:e0+ct}
\end{equation}
where $\omega$ is the pion energy in the center-of-mass;
these are significantly changed from one fit to another.
In this analysis one has to keep in mind that the $E_{0+}$ multipole is a problematic one in HBCHPT due to its slow convergence \cite{CHPT91,CHPT96,CHPT01,AB-Delta}.
In principle these can be computed within Lattice QCD, and hence the extraction from data through
HBCHPT becomes very important for the correct understanding of effective field theory.

\begin{figure}
\rotatebox{-90}{\scalebox{0.33}[0.33]{\includegraphics{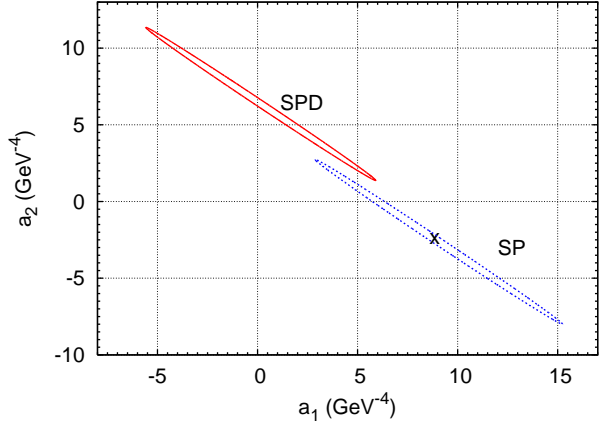}}}
\caption{(Color online)
Correlation plot for $a_1$ and $a_2$ given the P-wave parameters. We display both the
SP (dotted) and SPD (solid) models. 
The cross is the value of $a_1$ and $a_2$ given by Bernard \textit{et al.}  \cite{CHPT01}.
For each model, the line is the 70\% confidence level.
Clearly seen are the strong correlation between the two parameters and the large
uncertainties in their values.} \label{fig:a1a2}
\end{figure}

As noticed by Bernard, Kaiser, and Mei\ss ner \cite{CHPT01}, $a_1$ and $a_2$ are two highly
correlated LECs, and their sum is a better quantity to use when analyzing data.
In Fig. \ref{fig:a1a2} we display the correlation plot for $a_1$ and $a_2$, where we can see that
the two parameters
are strongly correlated and that the error bars are extremely large.

However, one can transform the counter-term from $(a_1, a_2)$ LECs to 
$(a_+=a_1+a_2,a_-=a_1-a_2)$ LECs, obtaining:
\begin{equation}
E_{0+}^{ct}= \frac{1}{2}e\omega M_\pi^2 \left[  a_+ \left( 1 + \frac{\omega^2}{M_\pi^2} \right)   
+ a_- \left( 1 - \frac{\omega^2}{M_\pi^2} \right)  \right] \: .
\end{equation}
In the near threshold region $\omega \simeq M_\pi$ and so
\begin{eqnarray}
1 + \frac{\omega^2}{M_\pi^2} &\simeq& 2 + \mathcal{O} \left( \frac{\omega}{M_\pi}  \right)\\
1 - \frac{\omega^2}{M_\pi^2}  &\simeq& 0 - \mathcal{O} \left( \frac{\omega}{M_\pi}  \right)
\end{eqnarray}
and
\begin{equation}
E_{0+}^{ct} \simeq e \omega M_\pi^2 \left[ a_+   + \frac{a_+-a_-}{2} \: \mathcal{O} \left( \frac{\omega}{M_\pi}  \right) \right] \: . \label{e0+exp}
\end{equation}
Hence, the leading order for the counter-term is the one associated with $a_+$,
namely  the parameter
that can be well established by fitting data.
At the 70\% confidence level we obtain for the SP and SPD models,
$a_+ =   6.48^{+0.82}_{-0.93}$ GeV$^{-4}$ and
$a_+ =   6.57^{+0.83}_{-0.92}$ GeV$^{-4}$, respectively.

\begin{figure}
\rotatebox{0}{\scalebox{0.45}[0.45]{\includegraphics{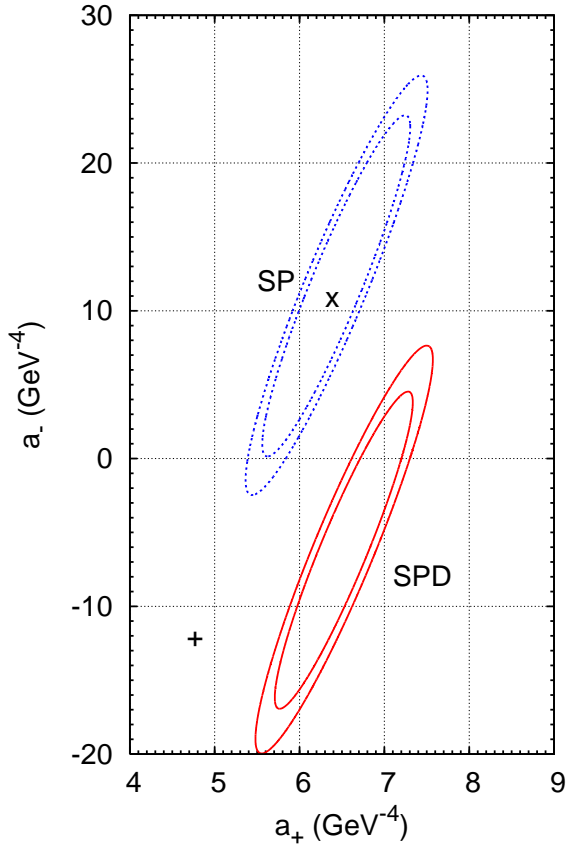}}}
\caption{(Color online) 
Correlation plot for $a_+$ and $a_-$ given the P-wave parameters. 
We display both the SP (dotted) and SPD (solid) models.
The cross is the value of $a_+$ and $a_-$ given by Bernard \textit{et al.}  \cite{CHPT01}
and the plus the value from Bernard \textit{et al.} \cite{VKM2005}.
For each model, the inner line is the 70\% confidence level and the outer one is the 90\% line.
Note that the different scales have been used for $a_+$ and $a_-$.} \label{fig:b1b2}
\end{figure}

In Fig. \ref{fig:b1b2} we show the correlation plot for $a_+$ and $a_-$, where we can see how the $a_+$ can be well determined from the fits and is not affected by the inclusion of the D waves.
On the other hand, $a_-$ is strongly affected by the inclusion of D waves, changing completely its value
and making the next order LEC $a_2=(a_+ - a_-)/2$ very difficult to assess, as  is shown in 
Figs. \ref{fig:a1a2} and \ref{fig:b1b2}. 

Another approach to extract the LECs is the one employed in \cite{VKM2005}, where the HBCHPT multipoles are fitted to the dispersion analysis of \cite{Pasquini} to compute
the Fubini-Furlan-Rossetti sum rule \cite{Fubini}.
This approach has the shortcoming that it relies on the extraction made through a dispersion analysis
with its own model dependencies and assumptions --- it relies on the phenomenological model for pion photoproduction MAID03 \cite{MAID} --- but has the advantages that HBCHPT
has better convergence inside the Mandelstam triangle and that D waves are incorporated
in the dispersion analysis.
Actually, in \cite{VKM2005} a truncated D-wave contribution was included in the fits
and the LECs obtained are closer to the ones we find. We display the 
$\left(a_+, a_- \right)$ pair from \cite{VKM2005} in Fig.  \ref{fig:b1b2} as a plus sign.

In summary,
contrary to what is customarily claimed in the literature, the S+P approximation is not sufficient
to obtain a complete description of the differential cross section in the near-threshold region
and to extract the electromagnetic multipoles reliably.
The inclusion of D waves does not affect the extraction of the 
P-wave multipoles, but makes a big
difference where the $E_{0+}$ extraction is concerned, especially in the energy region above the
unitary cusp.
The absence of D waves in the analysis compromises its extraction and the associated
LECs of the $E_{0+}$ counter-term.
The current experimental information is not yet accurate enough to pin down the $a_1$ and $a_2$ LECs,
but is capable of yielding a reasonable value for $a_+$, allowing one to extract a reliable  $E_{0+}$ at threshold to the order $q^4$ in HBCHPT. 
However, due to the slow convergence of the $E_{0+}$ 
multipole and the important effects of D waves and interferences one might
wonder if the $q^4$ calculation
is accurate enough and if a higher order computation may be necessary. 
The contribution of the D waves should be important in the interpretation of the new, more accurate,  pion photoproduction data which was recently obtained by the Crystal Ball Collaboration at Mainz 
\cite{Hornidge}, as well as other future photo-pion production experiments with polarization degrees of freedom \cite{AB-review,Bernstein}.

\begin{ack}
This research was supported in part (CF-R) by 
\textquotedblleft Programa Nacional de Movilidad de Recursos Humanos del Plan Nacional I+D+I 2008-2011\textquotedblright
of Ministerio de Ciencia e Innovaci\'on (Spain).
This work was also supported in part (AMB and TWD)
by the US Department of Energy under contract No. DE-FG02-94ER40818. 
\end{ack}

\end{document}